\documentclass[aps,twocolumn,preprintnumbers,amsart]{revtex4}
\pdfoutput=1
\usepackage{graphicx,graphics,epsfig,subfigure,times,bm,bbm,amssymb,amsmath,amsfonts,amsthm,mathrsfs,MnSymbol}
\usepackage{gensymb}
\usepackage[matrix,frame,arrow]{xypic}
\usepackage[pdfstartview=FitH]{hyperref}

\usepackage{subfigure}
\hypersetup{
    colorlinks=true,       
    linkcolor=red,          
   citecolor=magenta,        
    filecolor=magenta,      
    urlcolor=cyan,           
    runcolor=cyan
}
\usepackage[pdftex]{color}
\usepackage{braket}
\usepackage{enumerate}
\usepackage[normalem]{ulem}
\usepackage[usenames,dvipsnames]{xcolor}
\usepackage{multirow}
\usepackage{mathtools}

\definecolor{orange}{rgb}{1,0.5,0}

\newcommand{\bes} {\begin{subequations}}
\newcommand{\ees} {\end{subequations}}
\newcommand{\bea} {\begin{eqnarray}}
\newcommand{\eea} {\end{eqnarray}}

\definecolor{gold}{rgb}{0.85,.66,0}

\newcommand{\beq}{\begin{equation}}

\newcommand{\eeq}{\end{equation}}

\newcommand{\ignore}[1]{}

\newcommand{\RomanNumeralCaps}[1]
    {\MakeUppercase{\romannumeral #1}}

\def\tr{\mathrm{Tr}}

\def\bf{\mathbf{f}}

\def\s{\sigma}

\def\P{\Pi}
\def\S{\Sigma}

\def\>{\rangle}
\def\<{\langle}

\def\s0{I}

\newcommand{\ig}[1]{}

\begin{document}
\title{Quantum repeaters based on two species trapped ions}
\author{Siddhartha Santra$^{1}$}
\author{Sreraman Muralidharan$^{1,2}$}
\author{Martin Lichtman$^{3}$}
\author{Liang Jiang$^{2,4}$}
\author{Christopher Monroe$^{5}$}
\author{Vladimir S Malinovsky$^{1}$}

\affiliation{$^1$ US Army research laboratory, Adelphi MD 20783}
\affiliation{$^2$ Yale Quantum Institute, Yale University, New Haven, Connecticut 06520, USA}
\affiliation{$^3$ Joint Quantum Institute and Department of Physics, University of Maryland, College Park, MD 20742, USA}
\affiliation{$^4$ Departments of Applied Physics and Physics, Yale University, New Haven, Connecticut 06520, USA}
\affiliation{$^5$ Joint Quantum Institute, Department of Physics and Joint Center for Quantum Information and Computer Science, University of Maryland, College Park, MD 20742, USA}
\date{\today}

\begin{abstract}
We examine the viability of quantum repeaters based on two-species trapped ion modules for long distance quantum key distribution. Repeater nodes comprised of ion-trap modules of co-trapped ions of distinct species are considered. The species used for communication qubits has excellent optical properties while the other longer lived species serves as a memory qubit in the modules.  Each module interacts with the network only via single photons emitted by the communication ions. Coherent Coulomb interaction between ions is utilized to transfer quantum information between the communication and memory ions and to achieve entanglement swapping between two memory ions. We describe simple modular quantum repeater architectures realizable with the ion-trap modules and numerically study the dependence of the quantum key distribution rate on various experimental parameters, including coupling efficiency, gate infidelity, operation time and length of the elementary links. Our analysis suggests crucial improvements necessary in a physical implementation for co-trapped two-species ions to be a competitive platform in long-distance quantum communication.
\end{abstract}
\maketitle

\section{Introduction}
\label{sec:intro}
Communication using information encoded into quantum states of single photons offers unconditional security since any interception can be detected. This is the primary motivation behind quantum communication and cryptography \cite{Deutsch96,Shor2000,LoHoi-KwongandChau1999,Scarani2009,Tittel2002}. More generally, distributed entangled quantum states are a fundamental resource for quantum enhanced interferometry \cite{santra_qtel}, metrology \cite{Komar2014} and computation \cite{Beals-dqc}. However, the attenuation in the optical fiber poses a significant challenge for the long-distance transmission of single photons~\cite{Takeoka2014,Pirandola2017}. Quantum repeaters \cite{Briegel1998,Dur99} solve the attenuation problem by dividing the total communication distance into shorter channels connected by intermediate nodes, where the photon loss is detected \cite{Munro10,Bratzik14,Guha2015,Munro2008, Jiang2008,Epping2016} and can be corrected by using an active mechanism \cite{Muralidharan2016,Munro2015,Glaudell2016, Muralidharan2014,Muralidharan2017,Munro2012a,Namiki2016,Bergmann2016}. In addition to photon loss errors, operation errors accumulate over the quantum channel that degrades the quality of the transmitted entangled state. Depending on the methods used to overcome photon loss and operation errors, quantum repeaters make different demands on the quantum hardware for their construction \cite{Muralidharan2016,Munro2015,Muralidharan2014}. 

Trapped ions offer one of the most mature technologies for scalable universal quantum computing \cite{sciencekim,Monroe2014a,Tan2015} and quantum networking  \cite{Sangouard2009,QN-DuanMonroe}. Construction of large-scale quantum networks require matter qubits with long lifetimes, fast quantum gates and measurement, efficient coupling to photons, and low error rates of quantum operations \cite{Tan2015,Monz2011,Wang2017a,Ballance2016,Kim2011, Stute2012,Vandevender2010,Ghadimi2017,Kasture2016,simon-BSM}. Such a combination of characteristics is achievable if different elements with the desired properties are brought together in one technological platform. This is one of the reasons for the growing interest in two-species trapped ions (TSTI), for example, the pairs of $^{9}$Be$^{+}$ - $^{25}$Mg$^+$ or $^{171}$Yb$^{+}$ - $^{138}$Ba$^+$ ions \cite{Tan2015, multi-iontrap-network}. Here a species of ions can be utilized as a communication qubit ($^{25}$Mg$^+$ or $^{138}$Ba$^+$) that can be efficiently entangled to photons and a different longer lived species of ions ($^{9}$Be$^{+}$ or $^{171}$Yb$^{+}$) that can be utilized as a quantum-memory qubit and for local processing. A high-fidelity transfer of quantum information between the two species of ions can also be achieved \cite{Tan2015}. Furthermore, due to transition-frequency difference the optical operations with the communication ions do not disturb the quantum memory ions even though they are only a few microns away. Despite such experimental advances an analysis of quantum key generation rates, among the most promising uses of a quantum network, using two-species trapped ions for quantum repeaters is lacking.

We describe in this paper a quantum repeater architecture based on modules of two species of trapped ions. We study the viability of using these modules as building blocks for the construction of long distance quantum repeaters by analyzing quantum key distribution rates. The architecture can potentially work for any pair of communication and memory ions. For the numerical results later in the paper, however, we consider the advances made with the $^{171}$Yb$^{+}$ - $^{138}$Ba$^+$ pair described in \cite{multi-iontrap-network} with the following features: excellent isolation between $^{171}$Yb$^{+}$ quantum memory ions and $^{138}$Ba$^+$ communication ions was achieved and a state transfer between them was demonstrated; Ba$^+$ coherence times of $100~\mu$s in unstabilized magnetic fields and a coherence time of $4$ ms with stabilized magnetic fields were measured; same-species two-qubit gates between $^{171}$Yb$^{+}$ ions with $98\%$ fidelity and cross-species two-qubit gates between a $^{171}$Yb$^{+}$ ion and $^{138}$Ba$^+$ ion with $75\%$ fidelty in $200~\mu$s were demonstrated; light collection efficiency from the $^{138}$Ba$^+$ ion of about $10\%$ and fiber coupling efficiency of $17\%$ in the absence of a cavity were reported. 
Our analysis suggests crucial improvements in the performance parameters that can maximize key generation rates. We focus, in this paper, on simple repeater architectures that comprise of a single TSTI module at every repeater station. Based on the number of communication ions per module we classify TSTI modules as Type \RomanNumeralCaps{1} and Type \RomanNumeralCaps{2}. We then describe the modular operations necessary for quantum networking and highlight the operational differences between repeaters based on the module types in Sec.~(\ref{sec:describe}). The dependence of quantum key generation rates on experimental parameters such as gate error rates, operation time and coupling efficiency to fiber is presented in Sec.~(\ref{sec:errors}). In the same section we further compare the secure key generation rate achievable using the two basis protocol in the modular repeaters to the rate of reverse coherent information (RCI) \cite{RCI-shapiro} and the Pirandola-Laurenza-Ottaviani-Banchi (PLOB) bound \cite{Pirandola2017} that provide theoretical upper and lower performance benchmarks. Finally, we discuss the current experimental benchmarks and suggest the required improvements along with a concluding discussion in Sec.~(\ref{sec:conc}).
 

\section{Architectures for quantum repeaters}
\label{sec:describe}
\begin{figure}[h]
\centering
\includegraphics[width=\columnwidth]{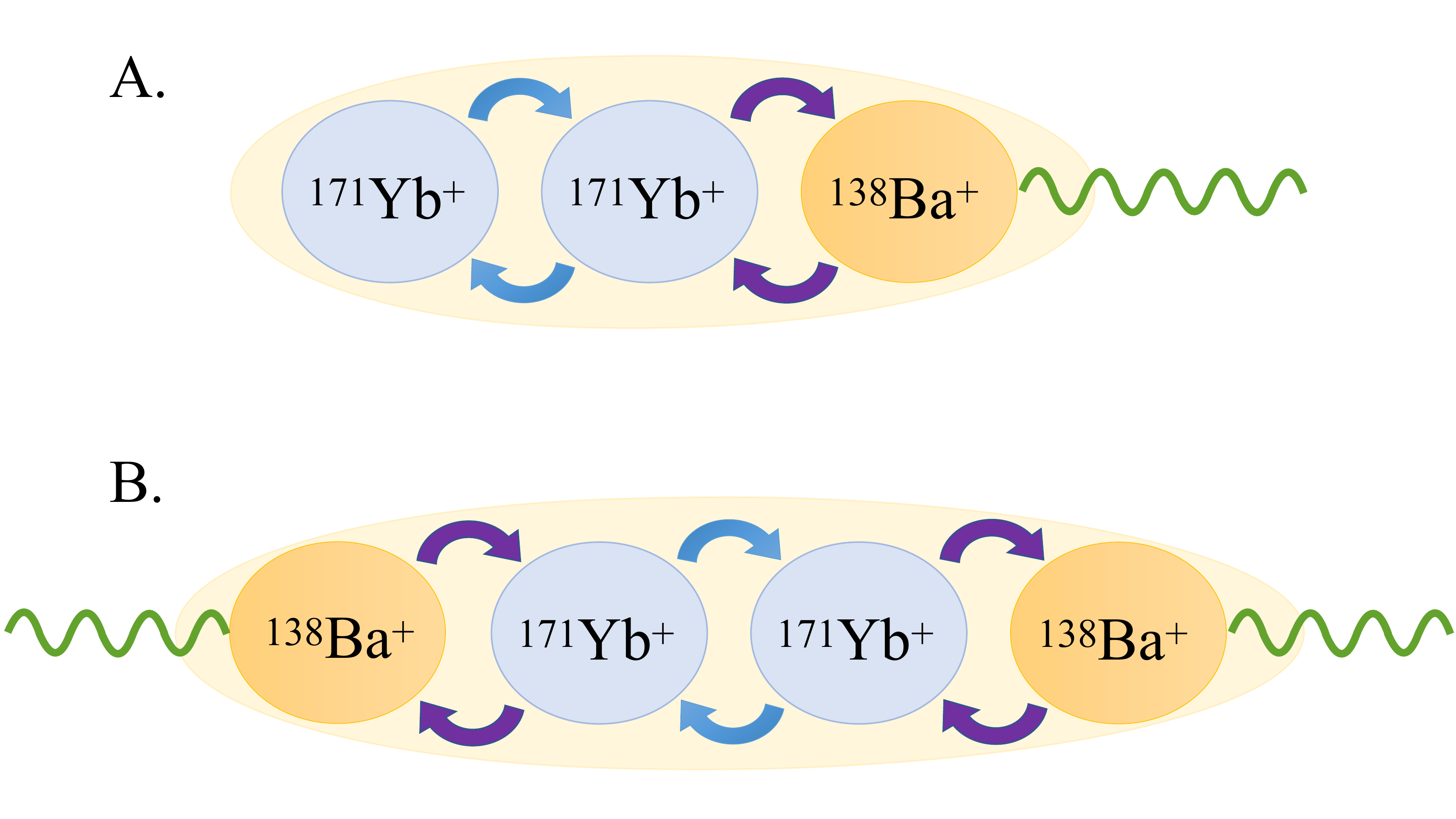} 
\caption{(Color online) Two species trapped ion modules with two memory ions (Yb$^+$) and a single communication ion (Ba$^+$) in panel (A) and with two memory ions (Yb$^+$) and two communication ions (Ba$^+$) in panel (B). The communication ions in any module interact with the quantum network via single photons (green wavy line). The communication and memory ions in the same module exchange quantum information via the quantum swap gate (purple arrows). Two memory ions in the same module undergo an entanglement swapping operation (blue arrows) to increase the range of entangled quantum states.}
\label{fig:twoions} 
\end{figure}

In this section we describe simple QR architectures feasible with TSTI modules. A single TSTI module consists of one or more  communication ions (Ba$^+$) and at least two memory ions (Yb$^+$), Figure \ref{fig:twoions}. We classify these modules, and the respective architectures which use these modules, by the number of their communication ions. The number of communication ions per module is a natural characteristic for classifying the modules and architectures because it determines the repeater protocol. Moreover, the number of communication ions per module also determines the technological complexity of the required hardware.  One or more modules can be stacked to form a single quantum repeater station. These modules interact with each other in a heralded manner only via photons emitted by the Ba$^+$ ions resulting in ion-ion entanglement. The entanglement of the communication ions is then coherently transferred to the memory ions in the respective modules using a quantum swap gate. A different entanglement swapping operation on distinct memory ions in the same module then yields entanglement length doubling. 

We focus, in the present work, on repeater architectures with a single TSTI module per repeater node and defer the discussion of architectures with multiple modules per repeater node to later work. Multiple TSTI modules per node permit encoding of the entangled quantum states but are challenging in the near term. Our goal in this paper is to identify technical improvements necessary in individual TSTI modules for viable quantum networking for which single modules per repeater node suffice. The elements of network operation described above categorize the simple architectures we consider in this paper into the second generation of quantum repeaters without encoding \cite{Muralidharan2014,Muralidharan2016,Munro2015}. In the following subsection~(\ref{subsec:elemops}) we describe the modular operations necessary for key generation and then in subsection~(\ref{subsec:arch}) we describe the simple repeater architectures we consider in the rest of the paper.
 
\subsection{Modular operations for quantum repeaters}
\label{subsec:elemops}

\subsubsection{Heralded entanglement generation between communication ions.}
\begin{figure}[h]
\centering
 \includegraphics[width=\columnwidth]{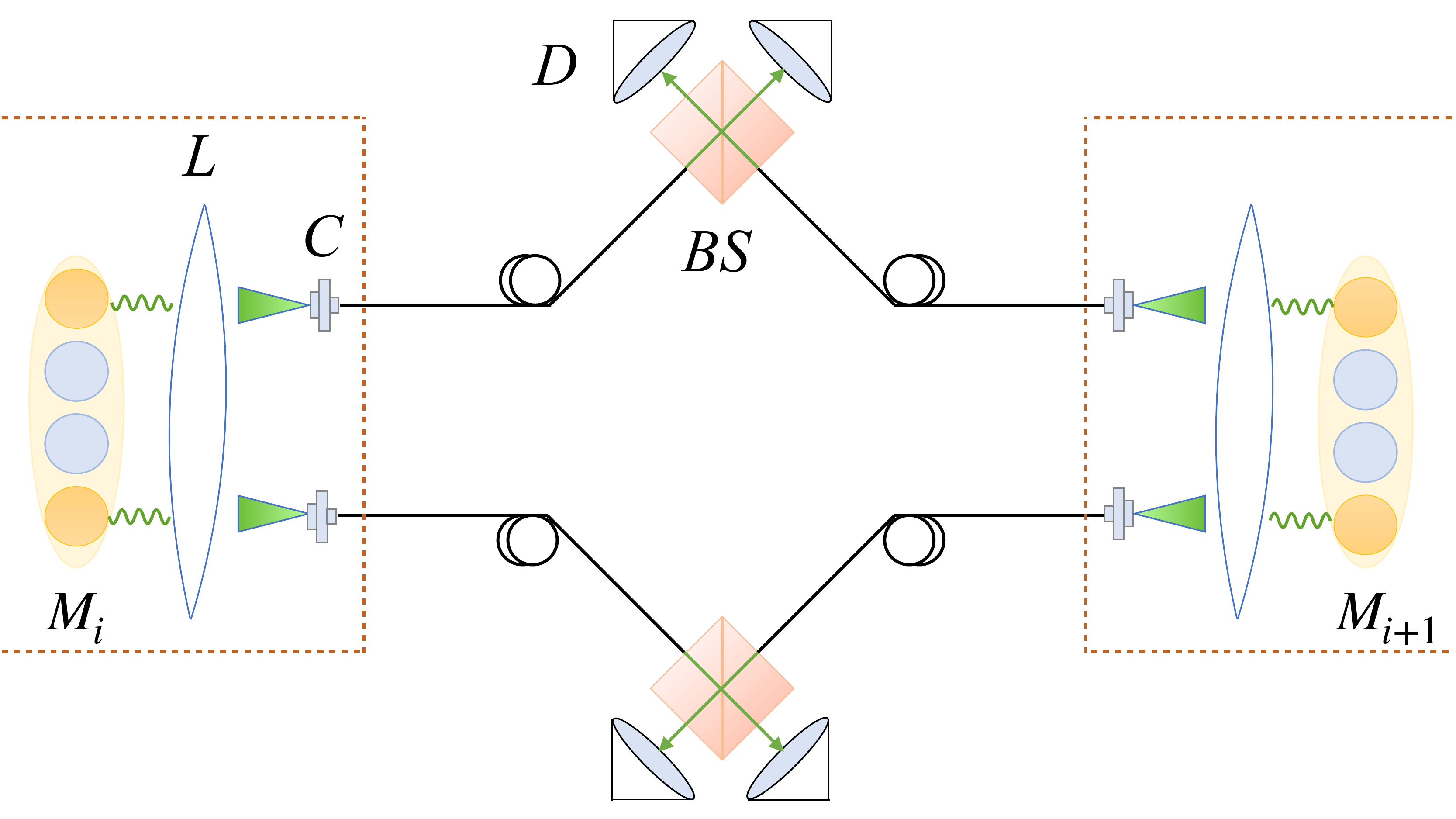} 
 \caption{(Color online) Setup for heralded entanglement generation between TSTI modules $M_i$ and $M_{i+1}$ at neighboring repeater stations. An optical setup, $L$, collects the photons emitted by the Ba$^+$ ions of a module. A fiber optic coupler, $C$, which includes a frequency conversion device channels the photons into an optical fiber. Photons from Ba$^+$ ions in two distinct modules travel through the fiber to the input ports of a beam splitter, $BS$. Their interference is monitored at the ouput ports of $BS$ by single photon detectors, $D$. Coincident detector clicks at the $BS$ outputs herald successful entanglement generation between the communication ions of two modules.}
\label{fig:2} 
\end{figure}

TSTI modules interact with the quantum network only via single photons emitted by the Ba$^+$ communication ions, Figure \ref{fig:2}. An optical setup collects the photons and channels it through a fiber coupler to an optical fiber. Appropriate frequency conversion to telecom wavelengths is performed for the fiber coupling. Photons from modules at neighboring repeater stations interfere at a beam splitter between the repeater stations. The coincident detection of two photons at the two outputs of the beam splitter then projects the state of the Ba$^+$ ions into an entangled state with a success probability of $\frac{1}{2}$ \cite{kirby-swapping}.

The success probability of generating entanglement between two Ba$^+$ ions located at neighboring stations in one trial is, 
$p=\frac{1}{2}\eta_{c}^{2}e^{-L_{0}/L_{att}}$, where $L_{0}$ is the spacing between neighboring repeater stations and $\eta_{c}$ is the coupling efficiency including emission, collection and coupling loses. Modules with multiple communication ions have a greater probability that at least one pair of ions across distinct modules are successfully entangled per trial. However, this is technically challenging since the ions within a module are separated only by a few microns. At such close separations, with a high probability a single photon emitted by a Ba$^+$ ion may be resonantly absorbed by a different Ba$^+$ ion in the same module instead of being collected by the optical setup. To mitigate such intra-modular absorption the Ba$^+$ ions must be distributed with maximum spacing among the Yb$^+$ ions along with spatially resolved photon collection.

\subsubsection{Quantum state swap between communication and memory ions.}
A quantum swap gate is used to transfer the state of the Ba$^+$ communication ion to an available Yb$^+$ memory ion in a TSTI module. This operation is required to avoid entanglement swapping directly using the entangled communication ions. By doing so, we circumvent the difficulty of measuring the Ba$^+$ ions \cite{multi-iontrap-network} and eliminate the effect of measurement errors.  For pure states, $\ket{\psi_A}$ and $\ket{\psi_B}$, of two Ba$^+$ ions the quantum swap gate, $\mathbb{S}$, exchanges the state of the two ions, i.e., $\mathbb{S}(\ket{\psi_A}\ket{\psi_B})\to\ket{\psi_B}\ket{\psi_A}$.
A pair of swap gates, $\mathbb{S}_{c_1m_1}$ and $\mathbb{S}_{c_2m_2}$, acting on two communication ions, $c_1$ and $c_2$, in an entangled state $\ket{\beta_{c_1c_2}}$ and two memory ions, $m_1$ and $m_2$, in unentangled pure states $\ket{\psi_{m_1}}$ and $\ket{\psi_{m_2}}$ transfer the entanglement to the memory ions,
\begin{align}
\mathbb{S}_{c_1m_1}\mathbb{S}_{c_2m_2}\ket{\psi_{m_1}}\ket{\beta_{c_1c_2}}\ket{\psi_{m_2}}=\ket{\psi_{c_1}}\ket{\beta_{m_1m_2}}\ket{\psi_{c_2}}.
\label{swapop}
\end{align}
This leaves the two Ba$^+$ ions in unentangled states $\ket{\psi_{c_1}}$ and $\ket{\psi_{c_2}}$. These communication ions may then be used for further rounds of heralded entanglement generation.
  
Gate operations on trapped ions are achieved via a sequence of laser pulses that controls dynamics of the ion wave functions. The desired swap gate in Eq. (\ref{swapop}) can be decomposed into a quantum circuit comprised of a set of native gates allowed by the specific experiment. Most relevant among the several physical implementations of such gates are the Cirac-Zoller (CZ) \cite{Cirac1995,Schmidt-Kaler2003} and the Molmer-Sorensen (MS) \cite{Molmer1999} gates. In the CZ-gate the motional mode of the ion crystal acts directly as a qubit transmitting quantum information. First, the internal state of one ion is mapped to the motion of an ion string followed by flipping the state of the target ion conditioned on the motion of the ion string. Finally the motion of the ion string is mapped back onto the original ion. On the other hand the main idea  behind of the MS gate is to drive collective spin flips of the involved ions. Ions can change their state only collectively and by choosing an appropriate amount of interaction time the desired two qubit unitary can be implemented. In addition, for MS gates individual addressing of ions is not required and the gates do not fail completely even if the temperature of the ion string is not absolute zero. We refer the interested reader to reference \cite{Haffner2008} for an excellent introduction to these mechanisms. Here we mention that our numerical analysis of key generation rates in later sections is independent of the particular gate mechanism used.

\subsubsection{Entanglement swapping between memory ions.}
\begin{figure}[ht]
\centering
\includegraphics[width=\columnwidth]{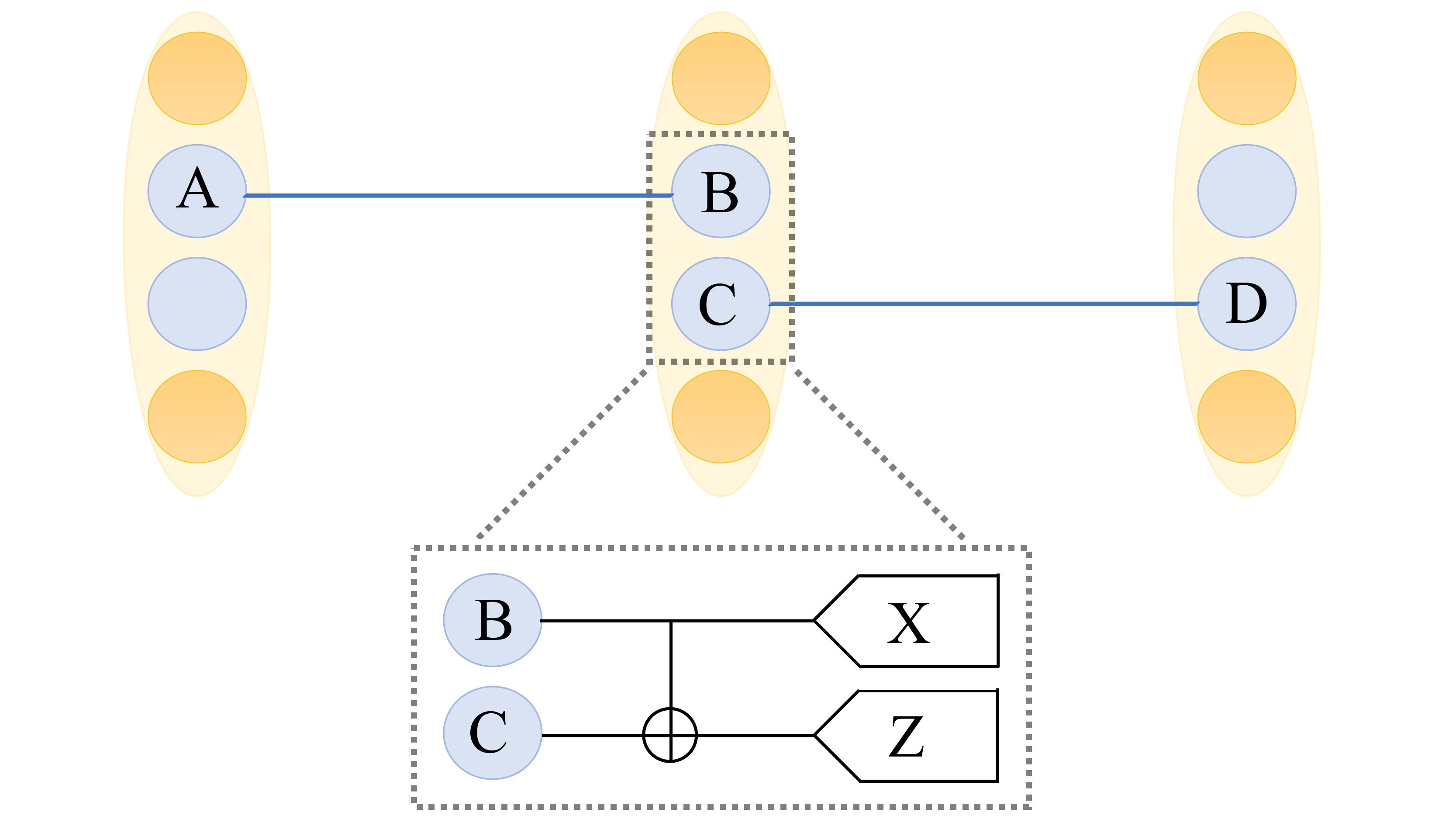}
\caption{(Color online) Entanglement swapping operation on memory ions at an intermediate repeater station. The two classical bits of the measurement result determine the one qubit unitary operations on the $A$ and $D$ ions to produce a Bell state.}
\label{fig:entswap} 
\end{figure}

A final step in our quantum repeater protocol is the entanglement swapping operation between Yb$^+$ memory ions within a  TSTI module. By connecting one memory ion each from two adjacent entangled states, this operation extends the physical distance of the entangled state. Entanglement swapping on two memory ions can be achieved using a quantum circuit consisting of a CNOT operation followed by a $X$ and $Z$ measurement as shown in Figure \ref{fig:entswap}. This procedure is equivalent to teleportation of the $B$ ion state to the $D$ ion. 
The entanglement swapping operation using coherent quantum gates on the memory ions is a deterministic process unlike the entanglement generation process which is probabilistic.

\subsection{Overview of architectures}
\label{subsec:arch}

We now describe quantum repeater architectures based on the module configurations and modular operations described above. Repeaters with two different kinds of modular configurations are considered: repeaters using Type \RomanNumeralCaps{1} modules with a single communication ion and Type \RomanNumeralCaps{2} modules with two or more communication ions. More communication ions per module greatly enhance the key generation rates but the technological requirements to increase the number of Ba$^+$ ions per trap, in a manner useful for quantum networking, also get significantly more complex. We assume in this paper that in both cases the repeater stations comprise of a single TSTI module.

\begin{figure}[h]
\centering
\includegraphics[width=\columnwidth]{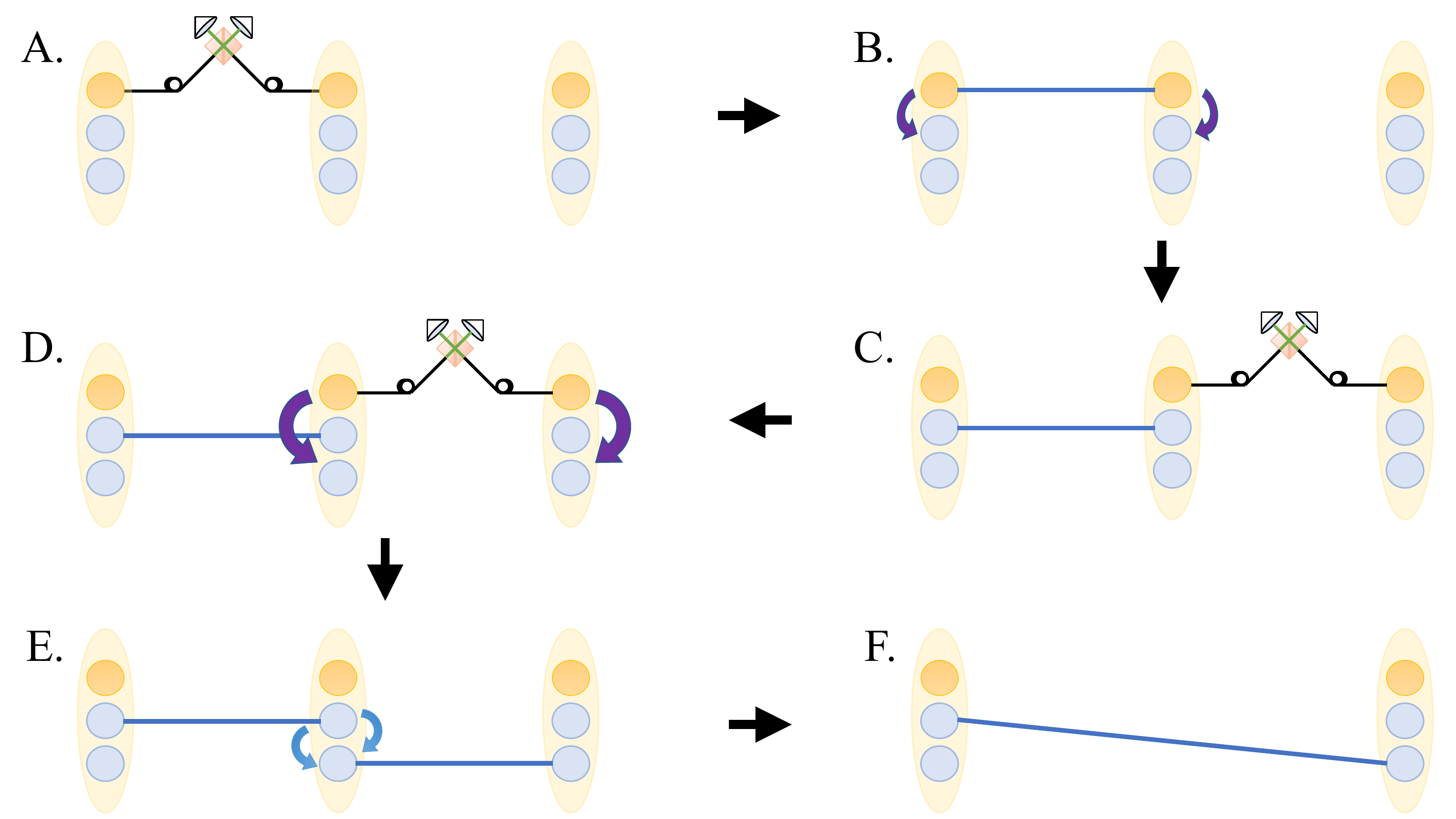}
\caption{(Color online) Quantum repeater architecture with Type \RomanNumeralCaps{1} modules with one communication ion and two memory ions. Three repeater stations with one module each are shown. Panels A-F show the sequence of modular operations required to obtain an entangled state across two repeater spacings. A) Heralded entanglement generation on the left. B) Pair of quantum swap gates to transfer the communication ion entanglement to the memory ions. C) Heralded entanglement generation on the right. D) Pair of quantum swap gates to transfer the communication ion entanglement to the other pair of memory ions. E) Entanglement swapping operation on the memory ions in the intermediate module. F) Creation of a long distance entangled pair.} 
\label{fig:sch2}
\end{figure}

In a quantum repeater with Type \RomanNumeralCaps{1} modules, entanglement is generated sequentially between the communication qubit of a module and another in its neighboring module - left followed by the right (or vice versa), Figure \ref{fig:sch2}. Starting with heralded entanglement generation attempts to the left of a repeater station, two-way classical communication between neighboring stations confirms its success. A pair of quantum swap gates then act on a communication-memory ion pair in each module to yield a pair of entangled memory ions across the modules. This process is then repeated for the module in the neighboring repeater station to the right. Once a pair of entangled memory ions are available both to the left and right of a module, an entanglement swapping operation is performed that extends the physical range of the entangled state. While technologically Type \RomanNumeralCaps{1} modules make modest demands, they are useful primarily for linear quantum repeaters.

\begin{figure}[h]
\centering
 \includegraphics[width=\columnwidth]{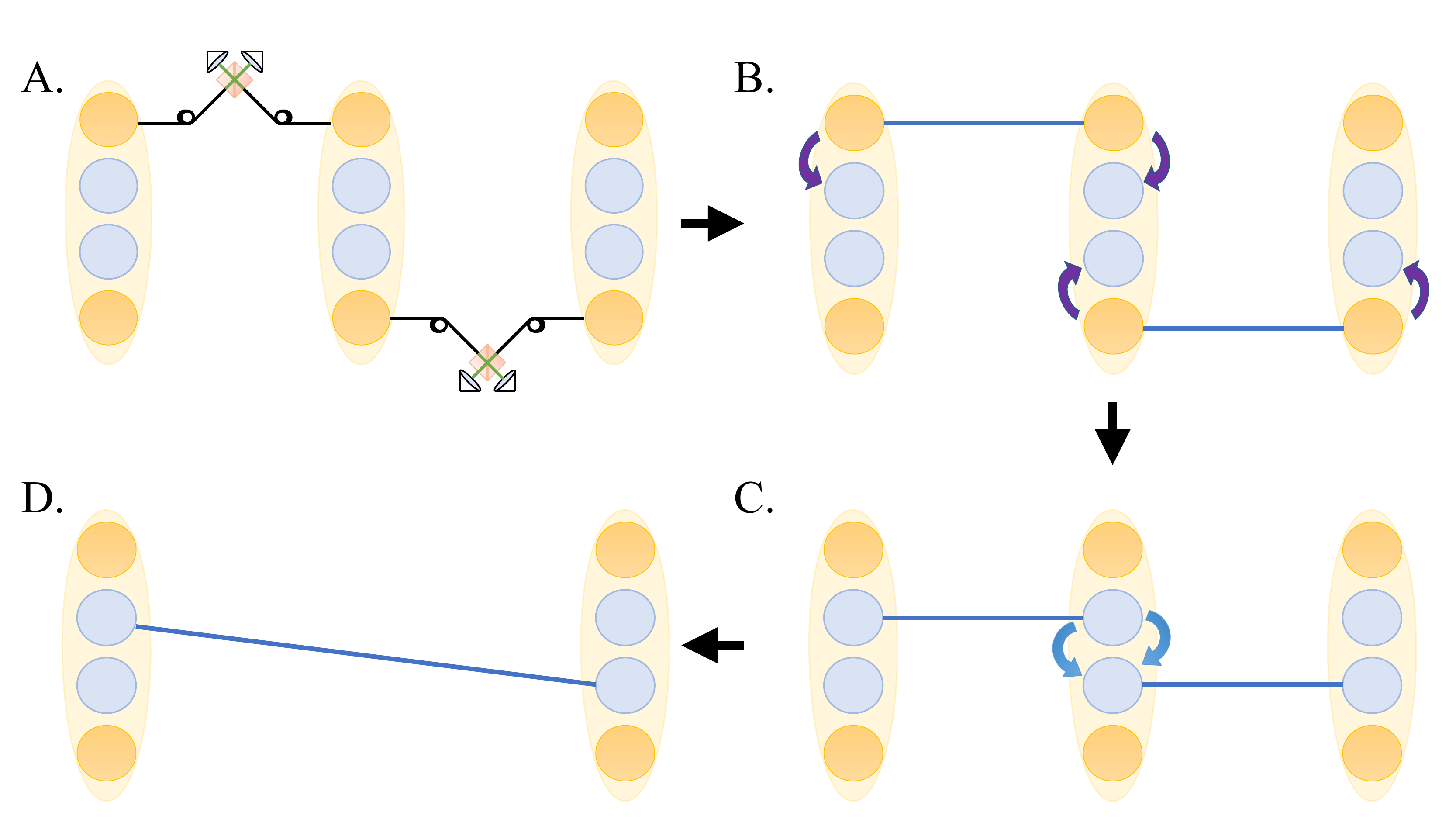} 
 \caption{(Color online) Quantum repeater architecture with Type \RomanNumeralCaps{2} modules with two communication and two memory ions. Three repeater stations with one module each are shown. Panels A-D show the sequence of modular operations required to obtain an entangled state across two repeater spacings. A) Heralded entanglement generation between communication qubits simultaneously in both directions. B) Pair of quantum swap gates to transfer the communication ion entanglement to the memory ions. C) Entanglement swapping of memory ions in the intermediate module. D) Creation of a long distance entangled pair.} 
\label{fig:sch1}
\end{figure}

In quantum repeaters with Type \RomanNumeralCaps{2} modules, the multiple communication ions in the same module allow simultaneous entanglement generation attempts with modules in neighboring repeater stations both to the left and right, Figure \ref{fig:sch1}. Type \RomanNumeralCaps{2} repeaters therefore obtain a significant boost in key generation rates. Moreover, such modules can be used in a true network configuration where the degree of a repeater node can be greater than two. However, Type \RomanNumeralCaps{2} modules are technologically more demanding. Photons from a communication ion in a module can be resonantly absorbed by another communication ion in the same module. Therefore, spatially addressable excitation and photon collection is required. Operationally, Type \RomanNumeralCaps{2} repeaters differ from Type \RomanNumeralCaps{1} repeaters in that the entanglement generation attempts do not need to be sequential. First, heralded entanglement generation with two-way classical communication confirms its success both to the left and right of a module. Then, a quantum swap gate transfers the entangled state of the communication ions to the memory ions of the respective modules. Finally, an entanglement swapping operation at intermediate stations is used to obtain a long distance Bell pair.

\section{Quantum key generation rates and error model for TSTI based repeaters}
\label{sec:errors}

Here we present the error model used to calculate the quantum key generation rate achievable with TSTI module based quantum repeaters. Errors contribute to the quantum bit error rate of the key generation protocol and lead to a degradation of key generation rates. Our error model accounts for two kinds of errors: a) Photon loss errors incurred during the entanglement generation step between communication ions and b) operation errors of the quantum swap gate and the entanglement swapping operation. These errors are estimated in terms of the parameters of gate time, gate error rates and coupling efficiency between communication ions and fiber in the Type \RomanNumeralCaps{1} and Type \RomanNumeralCaps{2} repeaters. In addition we assume that the initial Bell states generated between the communication ions of two neighboring modules have a nonzero infidelity to account for experimental imperfections.

\subsection{Error Model}
\label{subsec:errormodel}
The first kind of error we account for is the photon loss error which occurs because of fiber attenuation and a finite coupling efficiency, $\eta_c$, between the communication ions and the fiber. This effect of this error is to reduce the success probability of entanglement generation between communication ions of two different modules given by,
\begin{align}
p=\frac{1}{2}\eta_c^2 e^{-L_0/L_{\text{att}}},
\end{align}
where $L_{\text{att}}=20$ km for conventional optical fibers. The factor of $\frac{1}{2}$ occurs because of linear optical Bell state measurement \cite{kirby-swapping}. Photon loss errors reduce the rate of obtaining remote entangled states and result in lower raw key generation rates.

Second, we consider, the operation errors in the quantum gates and measurements used at the intermediate stations in the repeater protocols. This kind of error degrades the quality of the long distance entangled pair generated and contributes to the  quantum bit error rate of the key generation protocol resulting in lower key generation rates. The quantum swap gate requires three Molmer-Sorensen type gates (or two if one of the qubits is in a prepared state). The error in the swap gate is therefore three  (or two) times that of the CNOT gate to first order. We assume that these gate errors are parameterized by the larger of the gate errors among the swap and CNOT gates. We denote the effective error of the quantum channel, $\mathcal{E}$, representing a pair of quantum swap gates that transfers the entangled state, $\rho_{c_1c_2}$, of two communication ions to two memory ions by $\epsilon_{g}$. This implies that the output of this channel, $\rho_{m_1m_2}=\mathcal{E}(\rho_{m_1}\rho_{c_1c_2}\rho_{m_2})=\tr_{c_1,c_2}\{\mathbb{S}_{c_1m_1}\mathbb{S}_{c_2m_2}(\rho_{m_1}\rho_{c_1c_2}\rho_{m_2})\mathbb{S}^\dagger_{c_1m_1}\mathbb{S}^\dagger_{c_2m_2}\}$, obtained after tracing out the state of the communication ions is given by,
\begin{eqnarray}
\rho_{m_1m_2}  =(1-\epsilon _{g}) \rho_{c_1c_2}+ \frac{\epsilon _{g}}{16}\sum_{k^{\prime }=0}^{3}\sum_{k=0}^{3}\sigma_{k^{\prime }}\sigma _{k}\rho_{c_1c_2}\sigma _{k}\sigma
_{k^{\prime }},
\end{eqnarray}%
where, $\left\{ \sigma _{k}\right\} _{k=1,2,3}=\left\{ X,Y,Z\right\}$, are Pauli matrices and $\sigma_0=I$ is the identity operator on a single qubit. Note that the entangled state $\rho_{c_1c_2}$ generated between communication ions $c_1,c_2$ may itself have a non-unit fidelity $F_0$ and is assumed to be of the form, 
\begin{small}
\begin{eqnarray}
\rho_{c_1c_2} = F_0 |\varphi^{+}\rangle \langle\varphi^{+}|+\frac{(1-F_0)}{3}( |\varphi^{-}\rangle \langle\varphi^{-}|+|\psi^{+}\rangle \langle\psi^{+}| +|\psi^{-}\rangle \langle\psi^{-}|),
\end{eqnarray}
\end{small}
where, $\ket{\phi^\pm}=(\ket{00}\pm\ket{11})/\sqrt{2}$ and $\ket{\psi^\pm}=(\ket{01}\pm\ket{10})/\sqrt{2}$ are the complete set of the four perfectly entangled Bell states of the two communication ions. After performing a pair of quantum swap gates between memory and communication ions, the probability that there is either a $X$ or a $Y$ or a $Z$ error in any one of the qubits in the Bell pair is given by $\frac{(1-F_0)}{3}+\frac{\epsilon_g}{4}$. Further, the fidelity of the entangled pairs of memory qubits $\rho_{m_1m_2}, \rho_{m_3m_4}$ reduces after performing entanglement swapping on qubits $m_2,m_3$. Entanglement swapping requires measurements in the $X$ and $Z$ basis for qubits $m_2$ and $m_3$ respectively. While, $X$ and $Y$ type errors are detected in $Z$ measurements, $Z$ and $Y$ type errors are detected in $X$ measurements. The effective error measured at each repeater station can be approximated as, 
\begin{equation}
\epsilon_{X/Z} \approx \epsilon_g + \frac{2}{3} (1-F_0)
\end{equation}
Taking into account odd number of errors in repeater stations, the total quantum bit error rate is given by
\cite{Muralidharan2016},
\begin{equation}
Q_{X/Z}(R)= \frac{1}{2}[1-{(1-2\epsilon_{X/Z})}^R],
\label{eq:qber}
\end{equation}
where $R=(L_{\textrm{tot}}/L_0-1)$ is the number of repeater stations. We can define an effective quantum bit error rate as $Q(R)=(Q_X(R)+Q_Z(R))/2$. We will use this expression for the calculation of key generation rates in the forthcoming section.

\subsection{Secret key rates}
\label{subsec:ratecalc}

We now calculate the key generation rates for Type \RomanNumeralCaps{1} and \RomanNumeralCaps{2} repeaters. A total repeater length of $L_{\text{tot}}$ between remote locations, $A$ and $B$, with intermediate repeater stations at a spacing of $L_{0}$ is considered. For the Type \RomanNumeralCaps{1} repeaters, 
the probability of having a remote entangled pair between $A$ and $B$ after $n_{\text{eg}}$ steps of heralded entanglement generation between all neighboring repeater stations is given by, 
\begin{equation}
P_{\text{success}} = {[1-{(1-p)}^{n_{\text{eg}}}]}^{L_{\text{tot}}/L_0}.
\end{equation}
The time taken for entanglement generation in one repeater segment is given by 
\begin{equation}
T = L_0\frac{3n_{\text{eg}}}{2c} + 2t_0,
\end{equation}
where $t_0$ is the gate (swap or CNOT) and measurement time. The raw key generation rate is given by,
\begin{equation}
R^{\mbox{Type \RomanNumeralCaps{1}}}_{\text{raw}} = \frac{P_{\text{success}}}{2T},
\end{equation}
where in the denominator the factor of $2$ is due to the sequential generation of entanglement to the left and right of a repeater station. For the second architecture with $m$ communication ions in every module, the raw
key generation rates is given by
\begin{equation}
R^{\mbox{Type \RomanNumeralCaps{2}}}_{\text{raw}} = \frac{[1-\mbox{Prob(m,0,$n_{\text{eg}}$)}]^{L_{\text{tot}}/L_0}}{T}
\end{equation}
where, $\mbox{Prob(m,i,$n_0$)} = {m \choose i}{[1-(1-p)^{n_0}]}^i {(1-p)}^{n_{0}(m-i)}$ is the probability to have $i$ elementary links with $m$ qubits with $n_0$ rounds of entanglement generation. Therefore, $[1-\mbox{Prob(m,0,$n_{\text{eg}}$)}]$ is the probability to have at least one entangled pair in any one repeater segment after $n_{\text{eg}}$ attempts. 
The secure key generation rates for the two basis protocol is given by \cite{Shor2000, Scarani2009},
\begin{equation}
R_{\text{sec}} = R_{\text{raw}}(1-2h(Q)),
\label{secrate}
\end{equation}
where $h(Q)=-Q\log_2(Q)-(1-Q)\log_2(1-Q)$, is the binary entropy of the quantum bit error rate $Q$ and $(1-2h(Q))$ is the secret key fraction of the protocol.

\subsection{Reverse Coherent Information and the PLOB bound}
The reverse coherent information (RCI) \cite{RCI-shapiro} and the Pirandola-Laurenza-Ottaviani-Banchi (PLOB) bound \cite{Pirandola2017} provide upper and lower performance benchmarks for the repeater types under the assumptions of the error model.

The RCI of the shared remote entangled state $\rho_{\text{AB}}$ between the end points of the repeater provides the maximum possible secure key generation rate over all possible key generation protocols. The remote entangled state has the form, 
\begin{small}
\begin{equation}
\rho_{\text{AB}}=(1-\frac{3Q(R)}{2})\ket{\phi^+}\bra{\phi^+}+\frac{Q(R)}{2}(\ket{\phi^-}\bra{\phi^-}+\ket{\psi^+}\bra{\psi^+}+\ket{\psi^-}\bra{\psi^-}),
\end{equation}
\end{small}
compatible with the error model for the gates, measurements, the form of the elementary entangled states and the quantum bit error rate given in \ref{eq:qber}. The RCI of $\rho_{\text{AB}}$ can be evaluated as, $I_R(\rho_{\text{AB}})=S(\rho_{\text{A}})-S(\rho_{\text{AB}})$, where $S(\rho_\text{X})$ is the Von Neumann entropy of the density matrix $\rho_\text{X}$ and $\rho_\text{A}=\tr_{\text{B}}(\rho_{\text{AB}})$. The RCI of $\rho_{\text{AB}}$ is therefore given by the expression,
\begin{equation}
I_R(\rho_{\text{AB}})=1+(1-\frac{3Q}{2})\log_2(1-\frac{3Q}{2})+\frac{3Q}{2}\log_2\frac{Q}{2},
\end{equation}
with $Q=Q(R)$. The rate of reverse coherent information distributed by the two types of repeaters is given by $R^{\mbox{Type \RomanNumeralCaps{1},\RomanNumeralCaps{2}}}_{\text{raw}}I_R(\rho_{\text{AB}})$. Note that the RCI, $I_R(Q)$ is always greater than (or equal to for $Q=0$) the secret key fraction, $(1-2h(Q))$, whenever the secret key fraction is greater than zero.

The PLOB bound yields the maximum secret key rate obtainable via direct transmission over lossy bosonic channels connecting the remote points $A$ and $B$. The maximum rate per channel use, $K$ can be expressed in terms of the transmittivity, $\eta$, of the channel as, $K(\eta)=-\log_2(1-\eta)$, where in our case, $\eta=\eta_c^2e^{-L/L_{\text{att}}}$, to account for the coupling of two photons to the fiber and the transmission probability thereafter. The temporal rate equivalent of the PLOB bound is therefore $R_{\text{source}}K(\eta)$, where $R_{\text{source}}$ is the rate at which we can generate entanglement between two remote communication ions in modules located at $A$ and $B$. Since the quantum gates in our numerical analysis have an operation time of $t_0=1~\mu$s we consider $R_{\text{source}}= 1~$MHz for comparing the PLOB rate with those obtained using the repeater protocols.

\subsection{Numerical results}
\label{subsec:numres}

We cast the numerical evaluation of secure key generation rates of Type \RomanNumeralCaps{1} and \RomanNumeralCaps{2} repeaters as a multi-parameter optimization problem. The secure key generation rate given in Eq.~(\ref{secrate}) depends on the following parameters: total repeater length, $L_{\text{tot}}$; repeater spacing, $L_0$; fiber coupling efficiency, $\eta_c$; quantum swap gate and CNOT gate error, $\epsilon_g$; gate and measurement time for quantum swap gate and CNOT gate, $t_0$; initial infidelity of Bell pair generation. A total repeater length of $L_{\text{tot}}=1000$ km and an initial infidelity of Bell pair generation of $10^{-4}$ is assumed. In all the plots we report the maximum obtainable secure key generation rates by optimizing over the number of entanglement generation steps, $n_{eg}$.

\begin{figure}[h]
\centering
\subfigure[]{
  \includegraphics[width=.5\columnwidth]{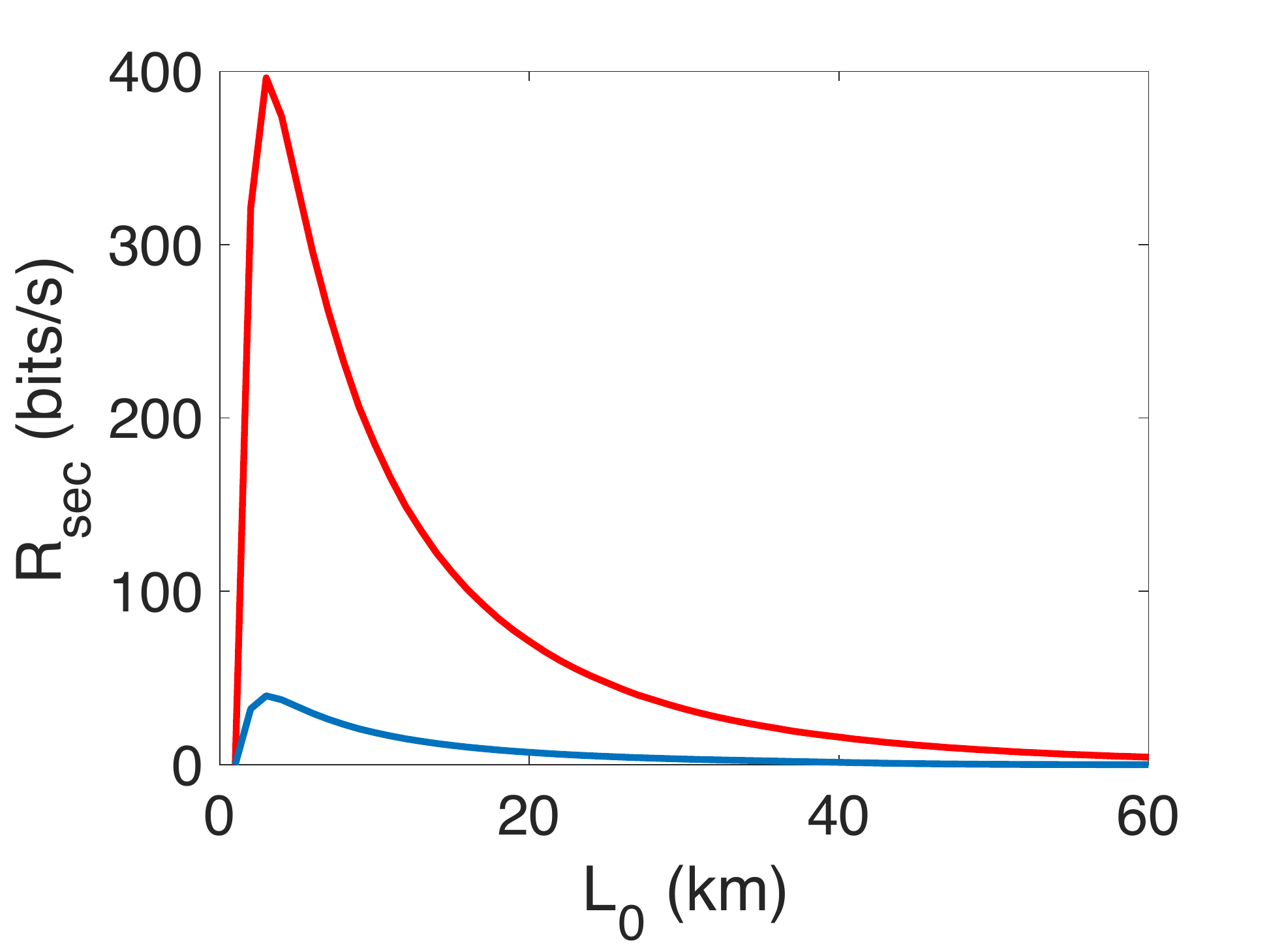}
   \label{fig:fig1}
 }\subfigure[]{
 \includegraphics[width=.5\columnwidth]{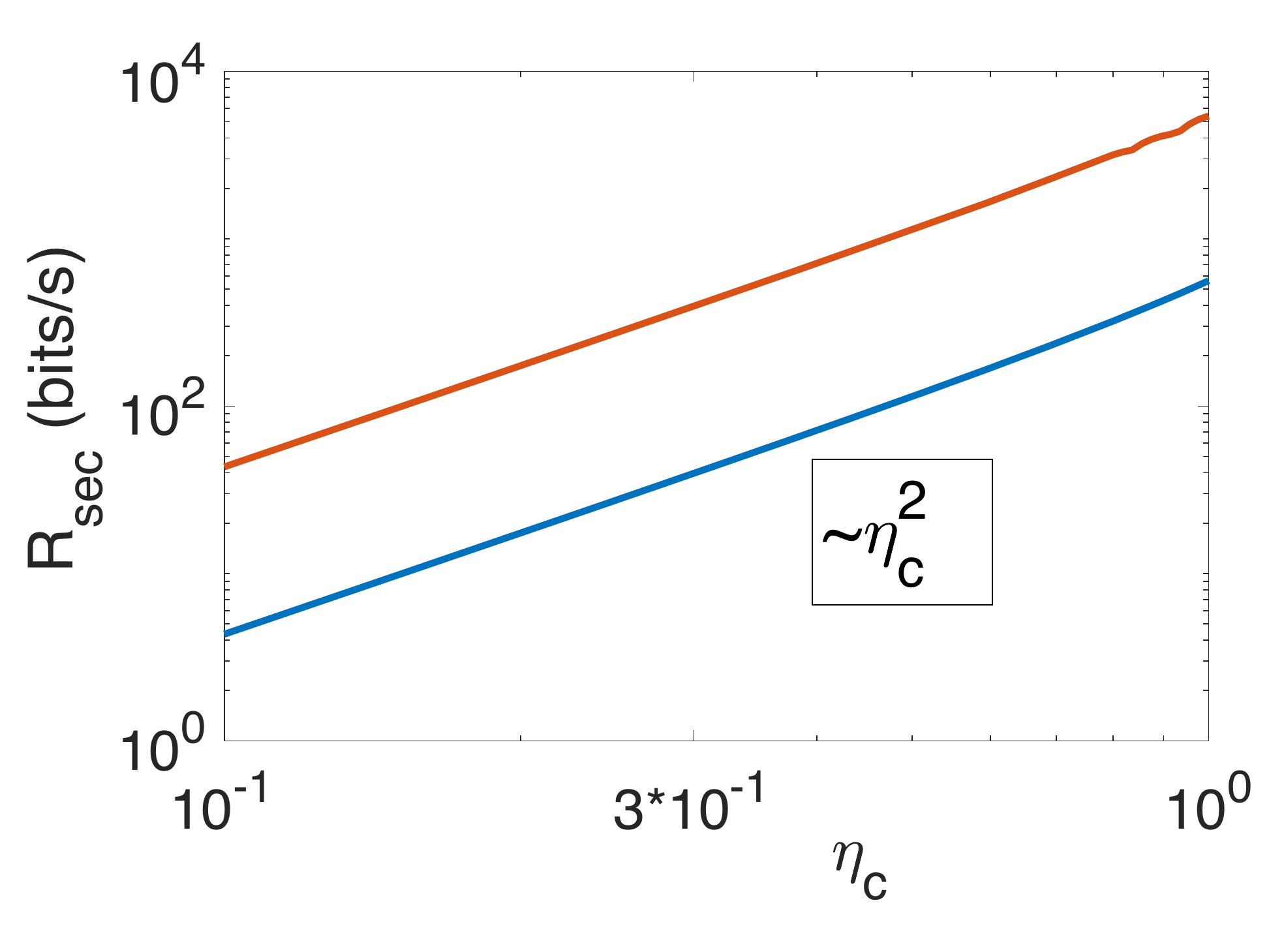}
   \label{fig:fig2}
 } 
\subfigure[]{
  \includegraphics[width=.5\columnwidth]{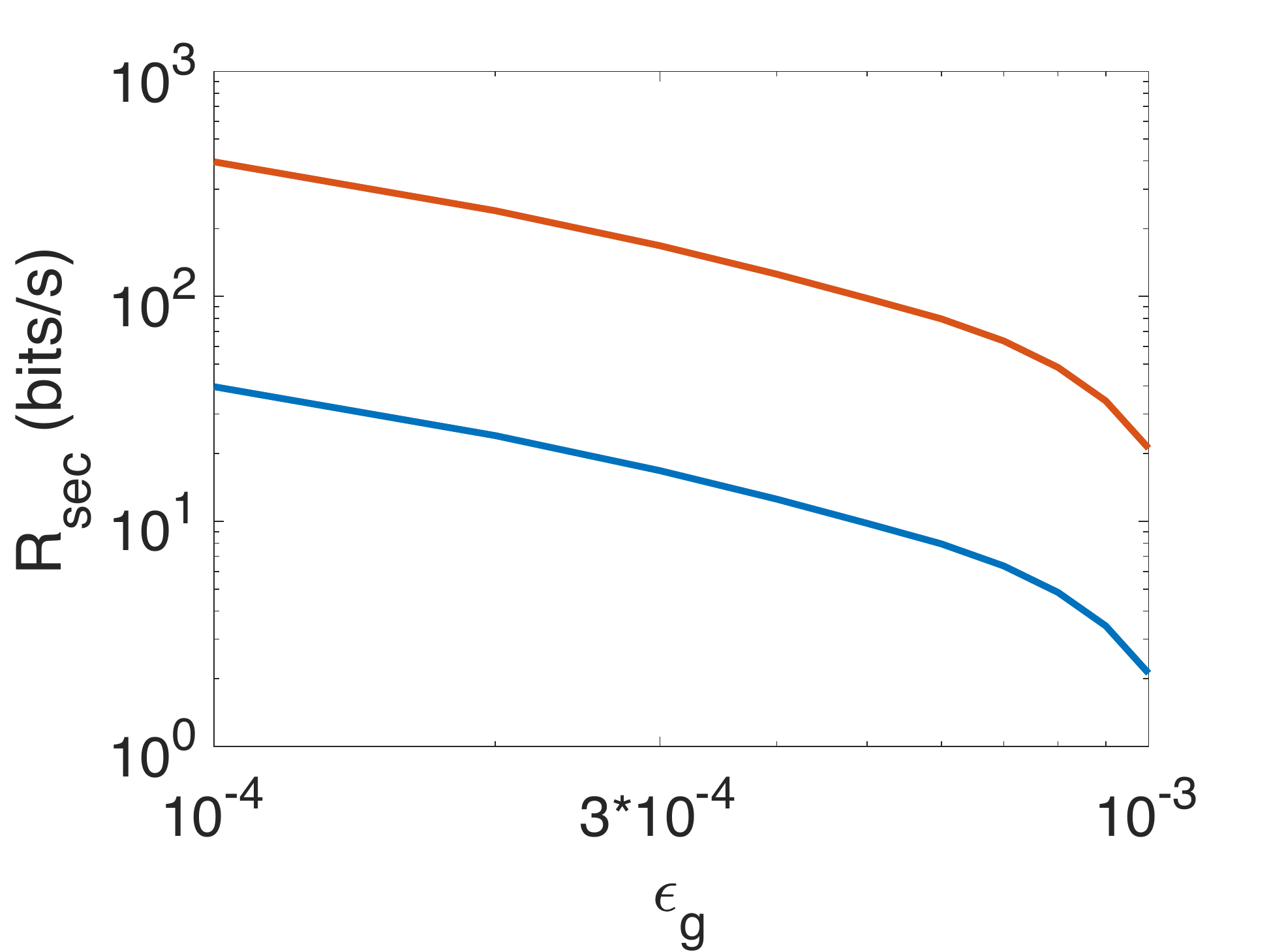}
   \label{fig:fig3}
 }\subfigure[]{
  \includegraphics[width=.5\columnwidth]{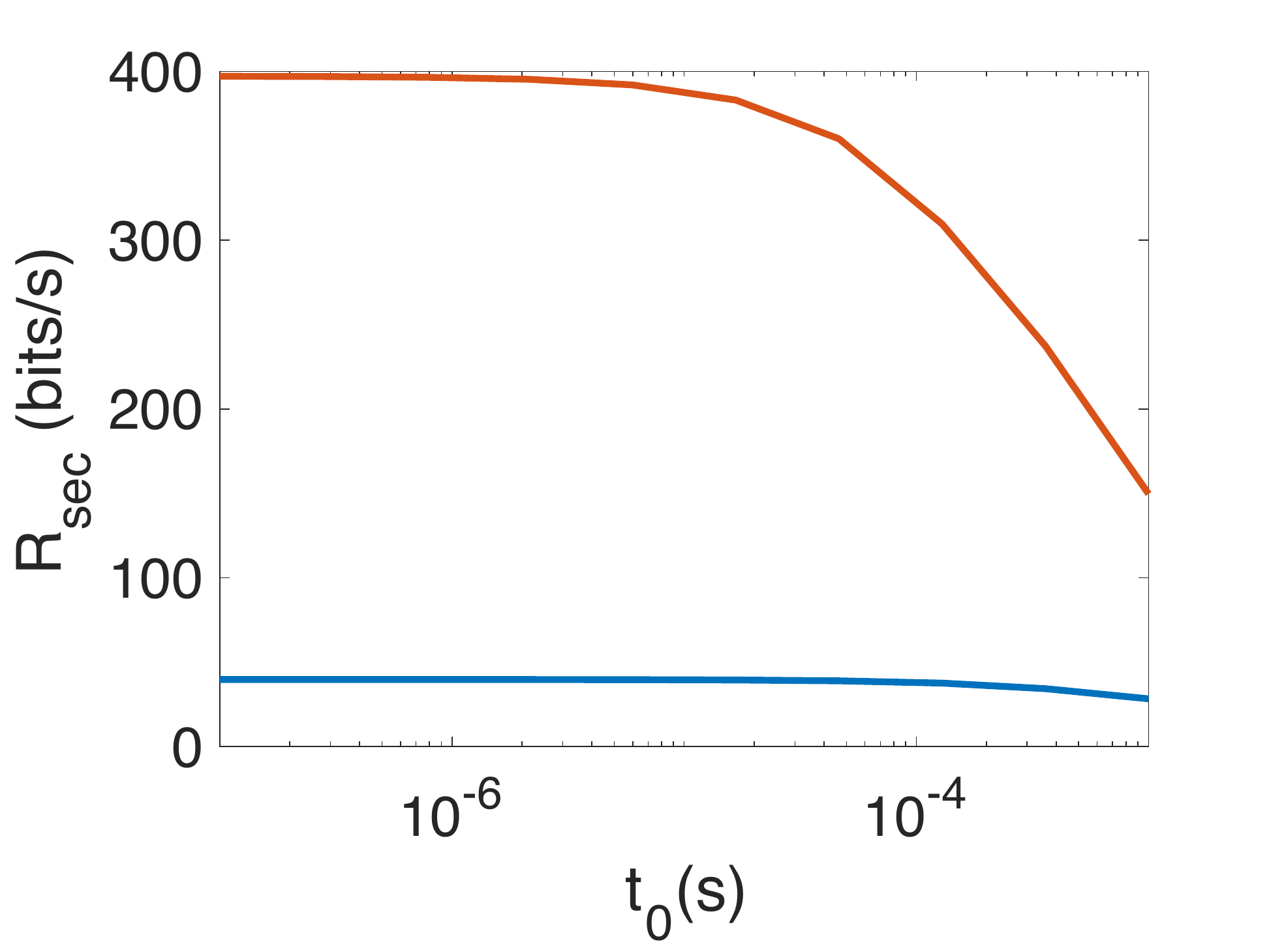}
   \label{fig:fig4}
 }
\caption{ (Color online) Variation of secure key generate rate with error model parameters for a total repeater length of $L_{\textrm{tot}}=1000$ km. Shown are the rates for Type \RomanNumeralCaps{1} repeater (blue curve) with 1 communication and 2 memory ions, Type \RomanNumeralCaps{2} repeater (red curve) with 10 communication and 2 memory ions. For all plots the number of entanglement generation attempts $n_{eg}$ has been optimized. Panels a-d show secure key generation rate with respect to a) repeater spacing, $L_0$, and $\eta_c=30\%$, $\epsilon_g = 10^{-4}$ and $t_0=1~\mu$s. b) coupling efficiency, $\eta_c$ and $\epsilon_g = 10^{-4}$ and $t_0=1~\mu$s; c) gate error rate, $\epsilon_g$, and $\eta_c=30\%$ and $t_o=1~\mu$s; d) operation time, $t_0$, and $\eta_c=30\%$ and $\epsilon_g = 10^{-4}$. }
\end{figure}

To start, we study the optimal repeater spacing - a crucial repeater design parameter that determines the key generation rate of Type \RomanNumeralCaps{1} and Type \RomanNumeralCaps{2} repeaters. On one hand, with small repeater spacings, i.e., many repeater stations, the operation errors dominate: multiple applications of the quantum swap gate and CNOT gate lowers the fidelity of the generated Bell pairs and increases the quantum bit error rate, $Q$. On the other hand, with large repeater spacing, i.e., small number of repeater stations, the loss errors dominate: a large number of entanglement generation steps are required leading to low secure key generation rates. This suggests that there exists an optimal repeater spacing for which the secure key generation rate is the highest at every value of $\eta_c$. We show the dependence of key generation rates with respect to repeater spacing in Figure \ref{fig:fig1}.

Next, we study the variation of key generation rate with respect to coupling efficiency, $\eta_c$, at the optimal repeater spacing as shown in Figure \ref{fig:fig2}. Reasonable key generation rates ($\sim$100 bits/s) are obtained for coupling efficiencies of about $70\%$ for Type \RomanNumeralCaps{1} repeaters with 1 communication and 2 memory ions (Type \RomanNumeralCaps{1} - 1C/2M). However, the rate achievable with Type \RomanNumeralCaps{2} repeaters with 10 communication and 2 memory ions (Type \RomanNumeralCaps{2} - 10C/2M) ions are about an order of magnitude higher. We note the sharp increase in the secure key generation rate by two orders of magnitude as $\eta_c$ increases to $100\%$ from $10\%$ and that $R_{sec}$ scales as ${\eta_c}^2$. This suggests that higher coupling efficiencies are necessary to achieve high secure key generation rates for all TSTI module based repeater architectures.
 
Further, the numerics accounts for the effect of operation errors by letting operation errors accumulate linearly with the number of repeater stations. As the operation errors adds up, the quantum bit error rate of the key generation protocol increases and the secure key generation rate reduces rapidly. The variation of key generation rates with respect to gate error rate is shown in Figure \ref{fig:fig3}. As it can be seen in the figure, the architectures tolerates operation errors up to $2.5\times10^{-3}$. Notice that reasonable secure key generation rates can still be achieved at error values well above the error correction threshold of $10^{-4}$ \cite{EC-threshold}.

Finally, the dependence of key generation rates on the gate operation time, $t_0$, is shown in Figure \ref{fig:fig4}. The quantum swap gate needed for transferring the quantum state from the communication to the memory ions and the CNOT gate required for entanglement swapping (including the $X$ and $Z$ measurements) take a finite amount of time which appears in the rate evaluation. Note that the secure key generation rate plateaus for faster gate times. Therefore, it is relatively insensitive to the gate operation time as long as gates are faster than a few microseconds.

\begin{figure}[h]
\centering
\subfigure[]{
  \includegraphics[width=.5\columnwidth]{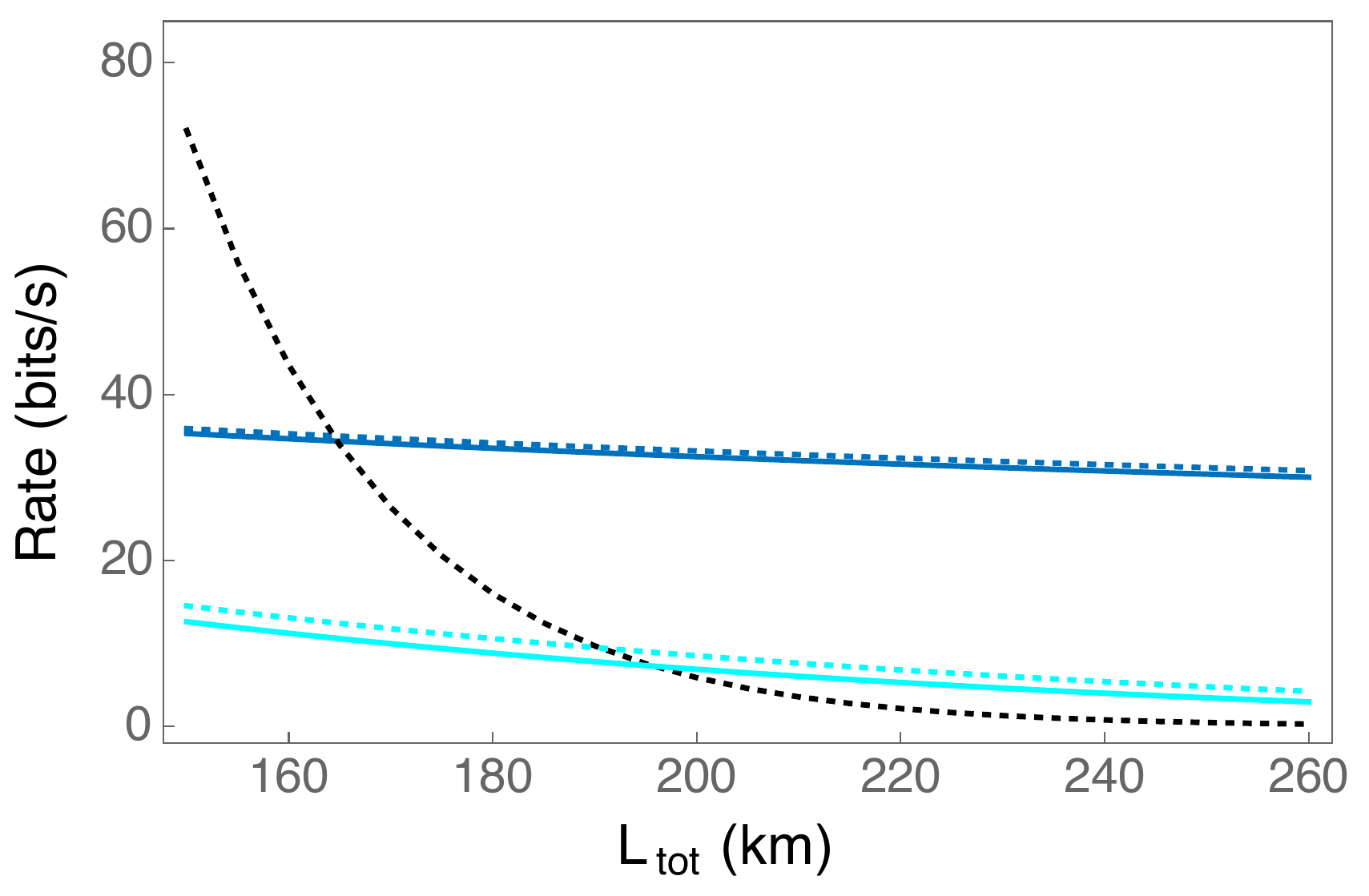}
   \label{fignew:fig1}
 }\subfigure[]{
 \includegraphics[width=.5\columnwidth]{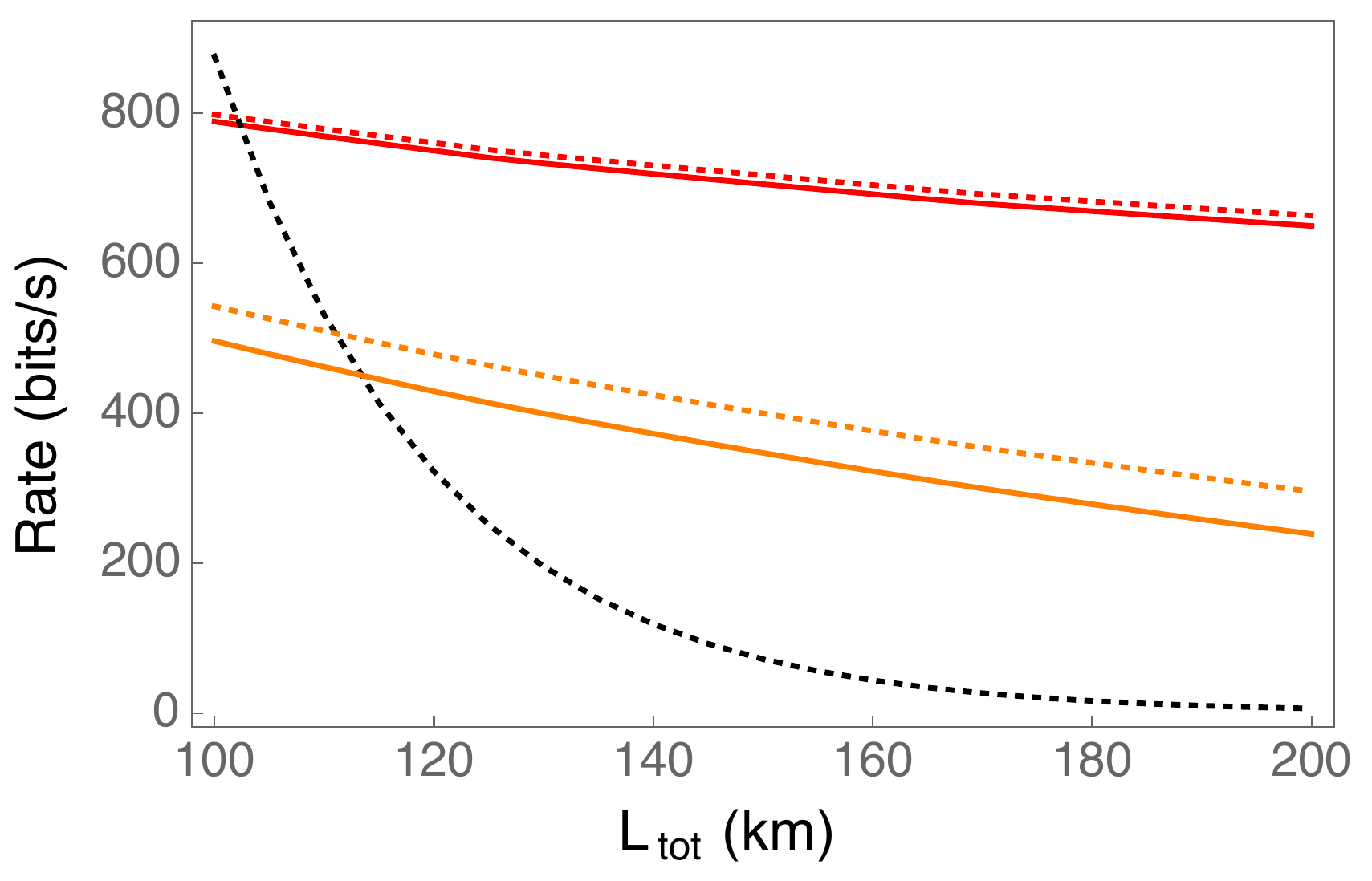}
   \label{fignew:fig2}
 } 
\caption{ (Color online) Secret key generation rate, Reverse coherent information rate and PLOB bound varying with total distance for Type \RomanNumeralCaps{1} (1C/2M) repeaters (left panel) and Type \RomanNumeralCaps{2} (10C/2M) repeaters (right panel). For both panels the repeater spacing is set at $L_0=5~$km and a fixed coupling efficiency of $\eta_c=30\%$ is assumed while the number of entanglement generation steps $n_{\text{eg}}$ is optimized. The noise parameters are $\epsilon_g=10^{-4}, F_0=(1-10^{-4})$ for the secret key rate (solid blue in panel (a), solid red in panel (b)) and RCI rate (dotted blue in panel (a), dotted red in panel (b)); $\epsilon_g=10^{-3}, F_0=(1-10^{-3})$ for the secret key rate (solid cyan in panel (a), solid orange in panel (b)) and RCI rate (dotted cyan in panel (a), dotted orange in panel (b)). The black dotted curve in both panels provides the PLOB bound.}
\end{figure}

We compare the performance of the two repeater types with the benchmarks provided by the RCI and the PLOB bound in Figure \ref{fignew:fig1} for Type \RomanNumeralCaps{1} and \ref{fignew:fig2} for Type \RomanNumeralCaps{2} repeaters. We observe that while the PLOB rate decreases exponentially with distance, the secure key rates obtained using both types of repeaters decrease polynomially. The secret key rate using direct transmission, i.e. the PLOB bound, falls below the rate for Type \RomanNumeralCaps{1} (1C/2M) repeaters for all total distances, $L_{\text{tot}}\gtrsim195$ km, for gate error and entanglement generation fidelities of $\epsilon_g=10^{-3}, F_0=(1-10^{-3})$ with a coupling efficiency of $\eta_c=30\%$. By improving these parameters, for example with $\epsilon_g=10^{-4}, F_0=(1-10^{-4})$, Type \RomanNumeralCaps{1} repeaters can beat direct transmission starting from an even shorter distance of $L_{\text{tot}}\gtrsim165$ km. The secret key generation rates for Type \RomanNumeralCaps{1} repeaters are close to the upper bound set by the rate of RCI of the remote entangled state at all total repeater distances. The difference between the RCI rate and the secure key rate is smaller for higher gate and entanglement generation fidelities. Similar behavior is observed for Type \RomanNumeralCaps{2} (10C/2M) repeaters with the main difference being that the PLOB bound can be beaten starting from even shorter distances. For gate error and entanglement generation fidelities of $\epsilon_g=10^{-3}, F_0=(1-10^{-3})$, Type \RomanNumeralCaps{2} repeaters surpass the PLOB bound for $L_{\text{tot}}\gtrsim115~$km while for $\epsilon_g=10^{-4}, F_0=(1-10^{-4})$ the bound is surpassed for $L_{\text{tot}}\gtrsim105~$km with fixed $\eta_c=30\%$. For the latter noise parameters the secure key rate of the two basis protocol is close to the RCI rate. However for the higher noise values the secure key rate is about $10\%$ lower than the RCI rate. 

\begin{table}[h]
 \begin{minipage}{.5\linewidth}
      \centering
        \begin{tabular}{|c|c|c|}
          \hline
Architecture&1C/2M&10C/2M \\ \hline 
$L_0$ (km) & 3 & 3 \\ \hline
$n_{\text{eg}}$ & 14 & 3 \\ \hline
Rate (bits/s) & 1125 & 5416 \\ \hline 
        \end{tabular}
    \end{minipage}%
    \begin{minipage}{.5\linewidth}
      \centering
      \begin{tabular}{|c|c|c|}
          \hline
Architecture&1C/2M&10C/2M \\ \hline 
$L_0$ (km) & 2.9 & 2.9 \\ \hline
$n_{\text{eg}}$ & 1825 & 365 \\ \hline
Rate (bits/s) & 4 & 43 \\ \hline 
        \end{tabular}
    \end{minipage} 
 \caption{Optimal repeater spacing, $L_0$, and number of entanglement generation steps, $n_{\text{eg}}$,
for Type \RomanNumeralCaps{1} (1C/2M) and  Type \RomanNumeralCaps{2} (10C/2M) quantum repeaters with $\eta_c=100\%$ (left table) and $\eta_c=10\%$ (right table). $L_{\text{tot}}=1000$ km, $\epsilon_g=10^{-4}$, $t_0=1~\mu$s and  $(1-F_0)=10^{-4}$.}
\label{tab:tab1}
\end{table}

To highlight the necessity of higher coupling efficiency, $\eta_c$, we present the optimized values of repeater spacing and the number of entanglement generation steps for $\eta_c = 100\%$ and $\eta_c = 10\%$ in Table \ref{tab:tab1}. The data reveals that while the optimal repeater spacing does not change much with the increase of $\eta_c$, the secure key generation rate shows dramatic improvement. For both Type \RomanNumeralCaps{1} and Type \RomanNumeralCaps{2} repeaters the secure key generation rate improves by a factor of two orders of magnitude. 

 \begin{figure}[h]
\centering
\subfigure[]{
  \includegraphics[width=.5\columnwidth]{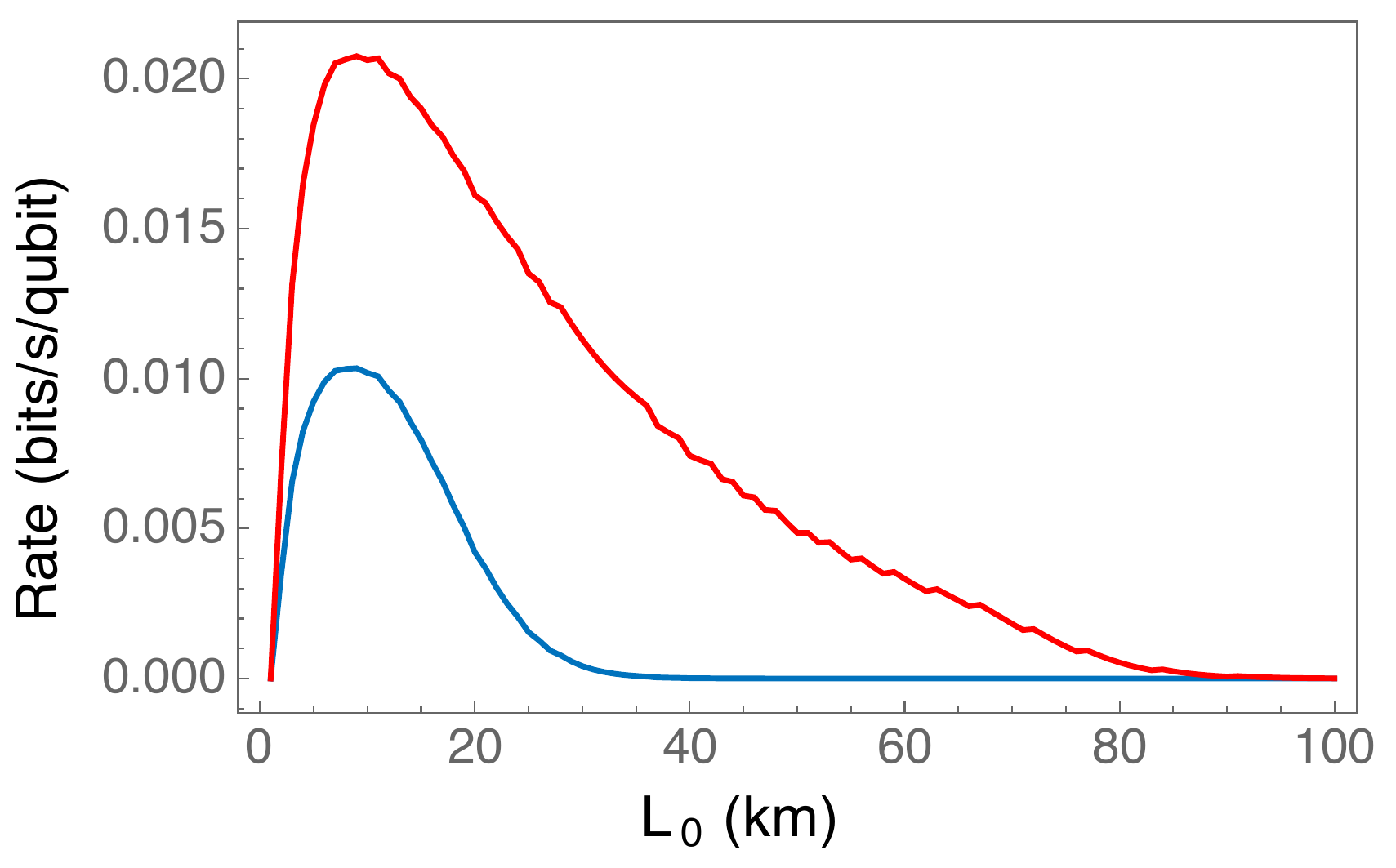}
   \label{fig:fig7}
 }\subfigure[]{
 \includegraphics[width=.5\columnwidth]{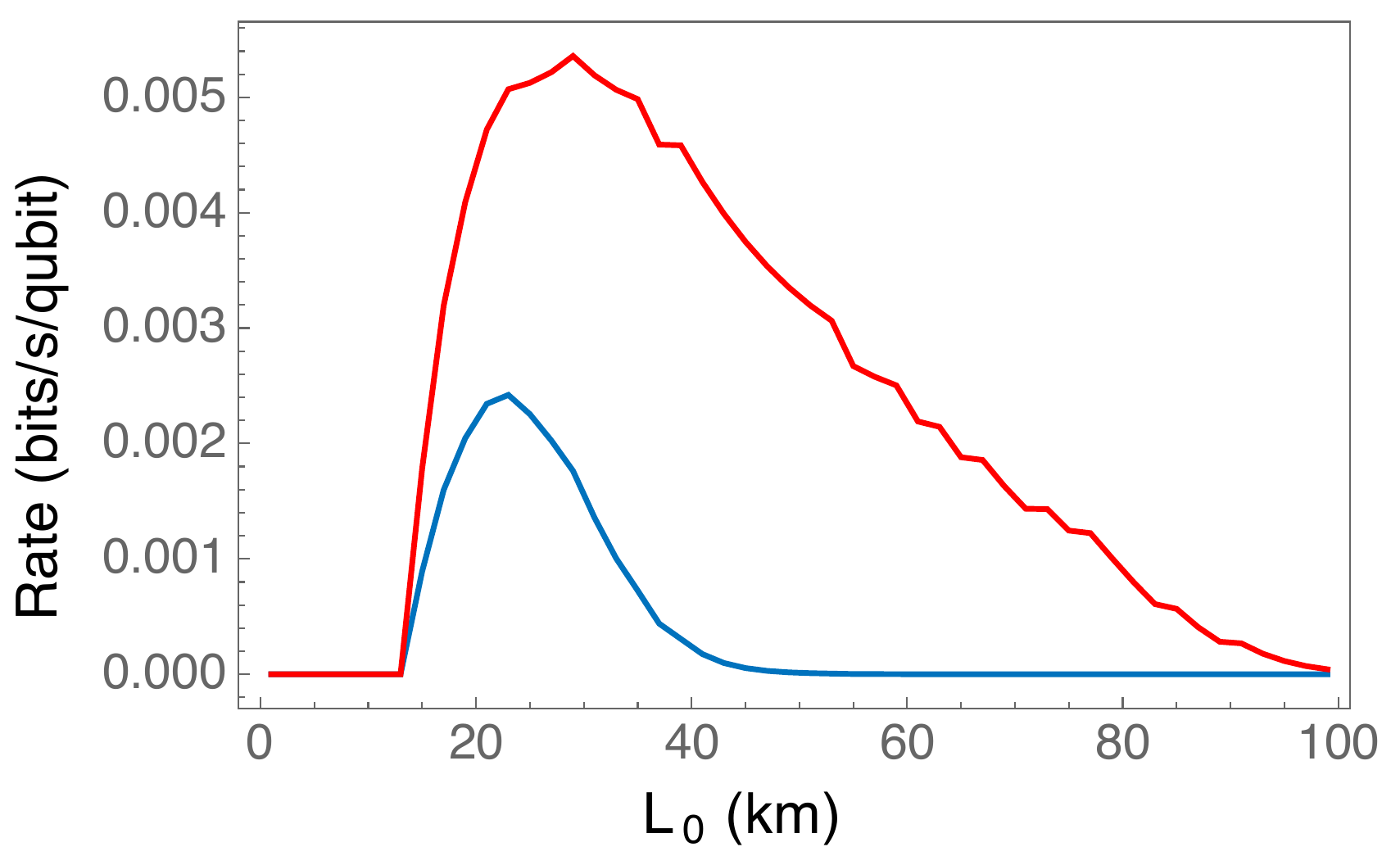}
   \label{fig:fig8}
 } 
\caption{(Color online) Variation of the secure key generation rate per deployed communication qubit with repeater spacing for Type \RomanNumeralCaps{1} (1C/2M, blue curve) and Type \RomanNumeralCaps{2} (10C/2M, red curve) repeaters. Panel (a) uses noise parameters of $\epsilon_g=10^{-4}, F_0=(1-10^{-4})$; Panel (b) uses noise parameters of $\epsilon_g=10^{-3}, F_0=(1-10^{-3})$. In both cases $\eta_c=10\%$, $t_0=1~\mu$s and $L_{\text{tot}}=1000$ km.}
\end{figure}

In a practical implementation the optimal repeater spacing needs to be determined by the overhead in the required physical resources. In that case the optimization metric is the secure key rate per deployed qubit in the entire chain of repeaters connecting the end points. The optimal repeater spacing maximizes this rate, $R_{\text{sec}}/(\text{number of repeater stations}\times\text{qubits per station})$, in units of  (bits s$^{-1}$/qubit) or conversely minimizes the cost coefficient \cite{Muralidharan2014}. In Figure \ref{fig:fig7} and \ref{fig:fig8} we plot the variation of the secure key generation rate per deployed qubit with repeater spacing at different noise levels for the two repeater types. For lower noise levels, $\epsilon_g=10^{-4}, F_0=(1-10^{-4})$, the optimal repeater spacing comes out to be $L_0\approx10~$km whereas for higher noise levels, $\epsilon_g=10^{-3}, F_0=(1-10^{-3})$, the optimal repeater spacing is about $L_0\approx20~$km for Type \RomanNumeralCaps{1} and $L_0\approx30~$km for Type \RomanNumeralCaps{2} repeaters. From the two plots, we also observe a minimum repeater spacing necessary for creating secure keys at a non-zero rate over $L_{\text{tot}}=1000$ km. For the lower noise level, the minimum repeater spacing is observed to be $L_0\gtrsim1.5$ km while for the higher noise level the minimum is $L_0\gtrsim15$ km. Repeaters spaced closer than these minimum values introduce too much operation error for the final remote state to be useful for secure key generation. Note that the optimal spacing based on optimization of physical resources is significantly different from the ones based only on the maximization of the secure key rate shown in Table \ref{tab:tab1}.

\begin{figure}[h]
\centering
\subfigure[]{
  \includegraphics[width=.5\columnwidth]{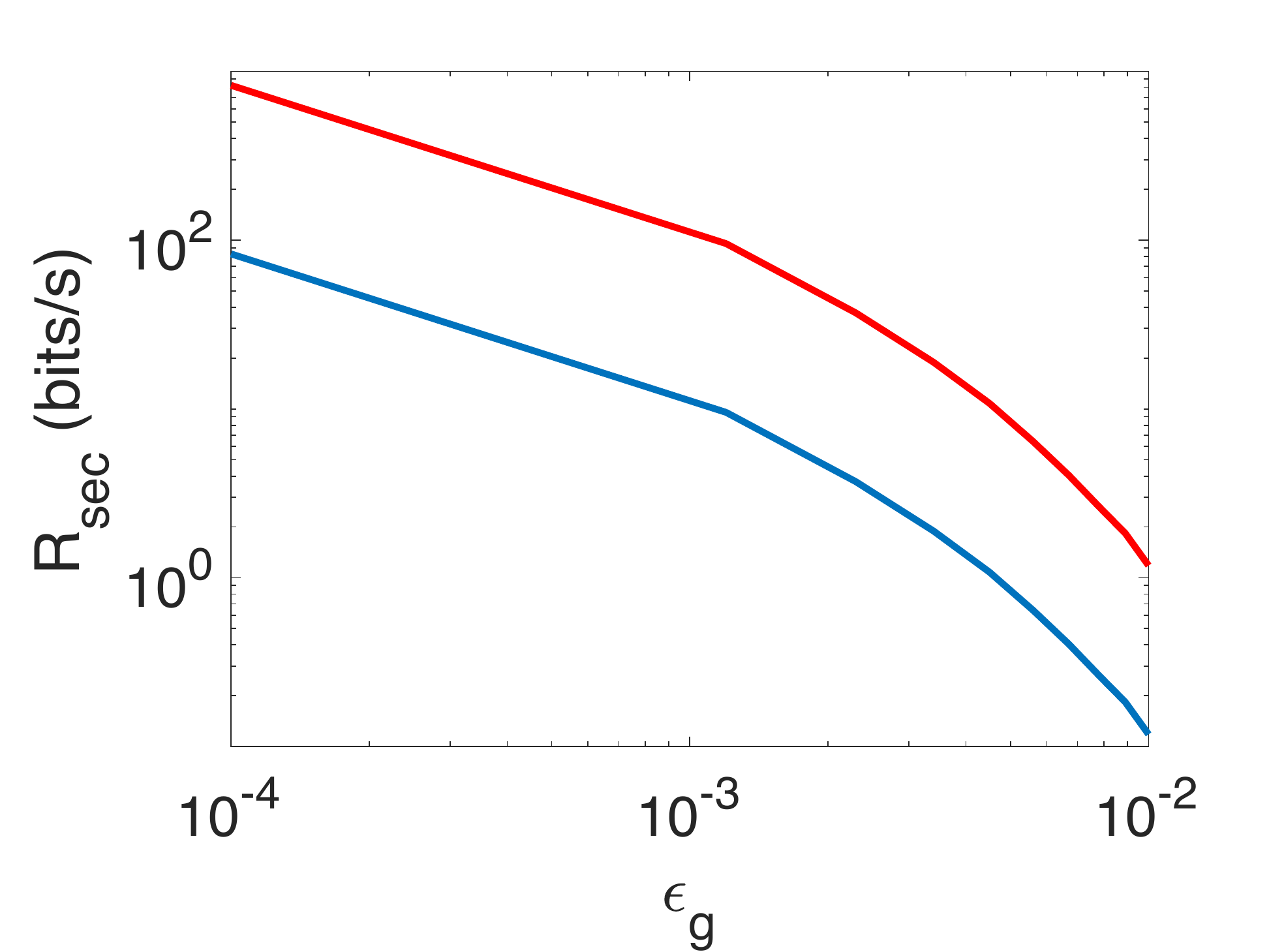}
   \label{fig:fig5}
 }\subfigure[]{
 \includegraphics[width=.5\columnwidth]{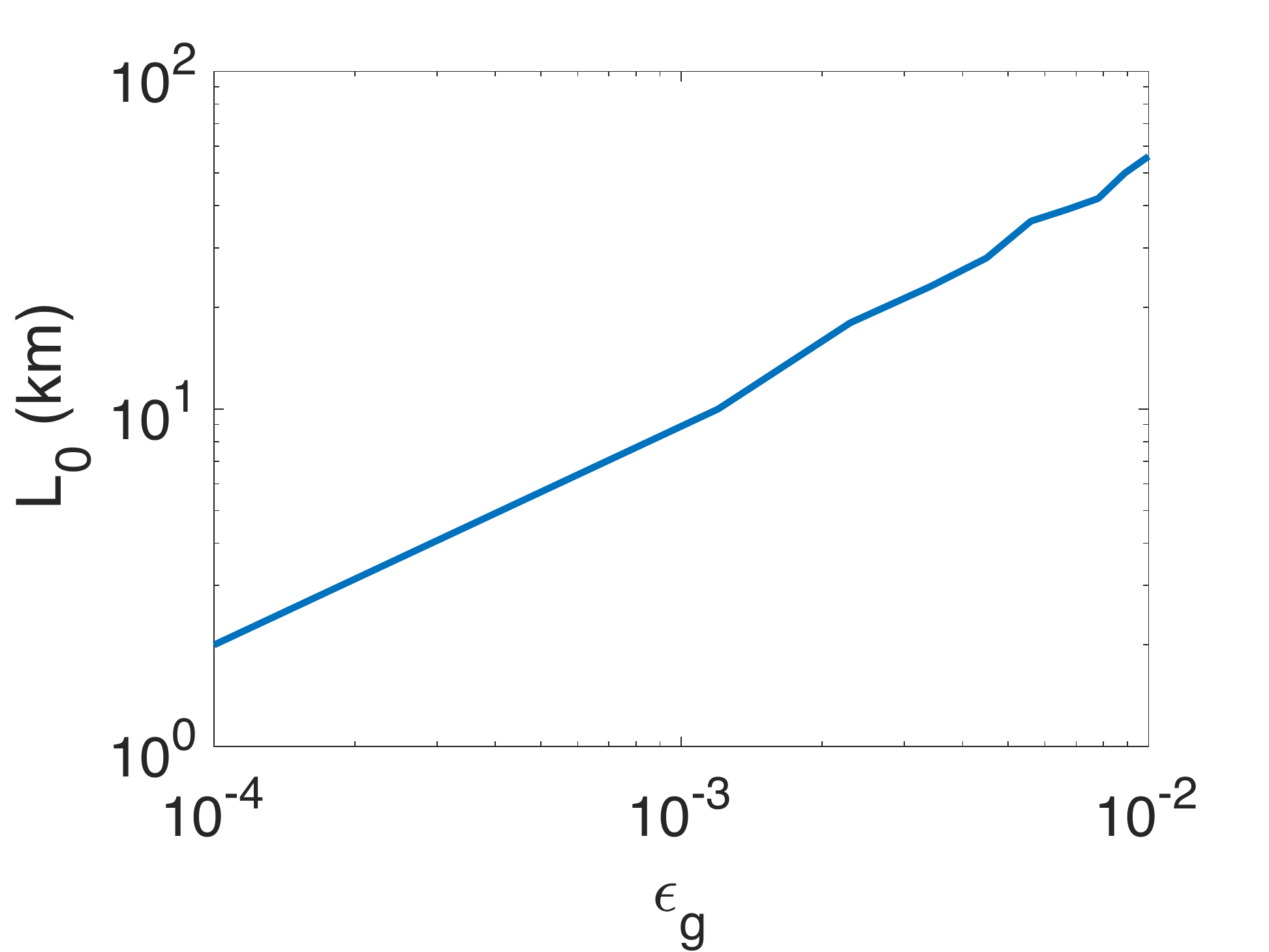}
   \label{fig:fig6}
 } 
\caption{(Color online) (a) Secure key generation rate with respect to gate error rates for Type \RomanNumeralCaps{1} (1C/2M, blue curve) and Type \RomanNumeralCaps{2} (10C/2M, red curve) repeaters with $\eta_c=30\%$, $t_0=1~\mu$s and $L_{\text{tot}}=500$ km. b) Optimal repeater spacing, $L_0$, of Type \RomanNumeralCaps{1} (1C/2M) repeaters with respect to operation error $\epsilon_g$ for $\eta_c=30\%$, $t_0=1~\mu$s and $L_{\text{tot}}=500$ km.}
\end{figure}

Our numerical results present an optimistic scenario since the experimental state-of-art fiber coupling efficiency is about $1.7\%$ between the Ba$^+$ communication ion and fiber in the absence of a cavity. The state-of-art gate error rate is about $2\%$ for two-qubit gates between ions of different species for a Be-Mg pair \cite{Tan2015}. At these error rates secure quantum key generation rates are highly suppressed for a total distance of $L_{\text{tot}}=1000$ km. However, these error rates for cross-species gates are sufficient to obtain reasonable key generation rates across a shorter total repeater length, $L_{\text{tot}}=500$ km, as shown in Figure \ref{fig:fig5} assuming perfect coupling between ion and fiber. For such large gate error rates while the key generation rates are low, the optimization picks higher repeater spacings $\sim 50-100$ km, as shown in Figure \ref{fig:fig6}, and a correspondingly higher number of entanglement generation attempts, $n_{\text{eg}}$.

To conclude this section, we remark that we studied the dependence of the secure key distribution rate on the repeater and the noise parameters as a multi-parameter optimization problem in Figures \ref{fig:fig1}-\ref{fig:fig4}. The secure key rates were compared with the RCI and PLOB bounds to obtain the distance for repeater advantage for different noise levels in the gates, measurements and fiducial entangled states in Figures \ref{fignew:fig1}-\ref{fignew:fig2}. In the latter comparison, we kept the repeater spacing fixed, however one can also keep the number of repeater stations fixed while their spacing changes for varying total repeater length. This results in an envelope of secure key generation rates which can be compared with the PLOB bound to obtain the distance beyond which the modular repeaters provide an advantage \cite{krovi}. Figures \ref{fig:fig7}-\ref{fig:fig8} present the optimal repeater spacing taking into account the overhead of required qubits which can be significantly different than the optimal spacing when only the secure key rate is maximized as in Table \ref{tab:tab1}. Figure \ref{fig:fig5}-\ref{fig:fig6} show that with state-of-art gate fidelities, reasonable secure key rates can be expected only over shorter distances of $\sim500$ km albeit with higher repeater spacing. Experimentally, to beat the secret key rates of direct transmission \cite{rozpedek-2018}, in addition to high gate and state fidelities, the fiber coupling efficiency needs to be improved to $\sim30\%$, perhaps by using adequate cavity QED effect, and a higher effective numerical aperture needs to be used to increase the collection probability to $\sim15\%$ \cite{Kim2011}. 
\section{Discussion and conclusion}
\label{sec:conc}

We studied quantum repeater architectures based on TSTI modules for their utility towards secure quantum key generation. Three elementary modular operations necessary for this application were identified: heralded entanglement generation between communication ions, quantum state swap between communication and memory ions and entanglement swapping between memory ions. We classified the modules into Types \RomanNumeralCaps{1} and \RomanNumeralCaps{2} based on the number of communication ions and associated technological complexity. Secure quantum key generation rates were calculated for repeater protocols based on the two basis protocol and modular operations assuming a general error model. The variation of the key generation rate with respect to repeater and error model parameters was studied and compared to the theoretical upper and lower bounds provided by the RCI and the PLOB bound.

Our analysis revealed that repeater designs using TSTI modules possess a combination of characteristics that is promising for applications such as quantum key distribution. The numerical analysis pointed out the improvements in the experimental parameters necessary to obtain reasonable secure key rates over long distances. Increasing the coupling efficiency was found to have the most beneficial effect on the rates while faster gate times resulted in more modest gains. The gate fidelities need to be considerably improved, to at least $99.9\%$ levels, especially for the quantum swap gate that acts on ions of two different species. Higher gate and elementary entangled state fidelities will lead to secure key generation rates close to the theoretical maximum rate given by the RCI rate and beat the direct transmission rate through lossy fiber optic channels faster. Finally, coherence time of the memory ion has to be further increased to allow for multiple rounds of entanglement generation \cite{Wang2017a} and potentially implement memory buffer-time optimized protocols for entanglement generation \cite{Santra_2019}. 

Besides trapped ions, there are several other competitive platforms in contention for the construction of quantum repeaters such as ensemble of atoms, Nitrogen-Vacancy (NV) centers and superconducting qubits. While an ensemble of atoms can be coupled to photons with a high efficiency of $85\%$ the only known approach to apply quantum gates is through linear optics which is limited by a finite success probability of $1/2$. This causes the key generation rates to drop exponentially with the number of entanglement swapping operations. NV centers can be coupled to photons with an efficiency of $15\%$, which is higher than the currently experimentally achievable coupling efficiency with communication ions. However, the spin-photon entanglement success probability is very low, about $10^{-6}$, for NV centers compared to ions which is about $0.07$. While superconducting qubits are another attractive candidate, the wavelength conversion efficiency between a single microwave photon and a telecom photon is expected to be very low. 

For future work, we will focus on quantum repeater architectures with more than one module per repeater station. These designs will harness the improvements in the experimental parameters of TSTI modules anticipated in the near term. Multiple interconnected modules, with higher coupling and low error gates, at the same repeater station will permit encoding of quantum states allowing robust key generation rates over long distances.

\section{Acknowledgements}
This work was supported in part by the Office of the Secretary of Defense, Quantum Science and Engineering Program.

S.M. acknowledge Linshu Li for discussions.

L.J. acknowledges support from the ARL-CDQI (W911NF-15-2-0067, W911NF-18-2-0237), NSF (EFMA-1640959), and the Packard Foundation (2013-39273). 

CM acknowledges support by the ARO with funds from the IARPA LogiQ program, the AFOSR project on Quantum Networks and MURI on Quantum Measurement and Verification, the ARL Center for Distributed Quantum Information, and the National Science Foundation Physics Frontier Center at JQI.

\section*{References}
\bibliographystyle{apsrev}

\end{document}